\documentclass[11pt]{article}
\usepackage{geometry}                
\usepackage{graphicx}
\usepackage{amssymb}
\usepackage{epstopdf}
\usepackage{amsmath, amssymb, graphics}
\newcommand{\mathsym}[1]{{}}

\title{Beam Fragmentation in Heavy Ion Collisions and its implication for RHIC triggers at low $\sqrt{s}$}
\author{Sebastian White\\
Physics Department, Brookhaven National Lab, Upton, NY 11973\\
Mark Strikman\\
Physics Department, Pennsylvania State University, University Park, PA 16802}
\date{October 16 '09}                                           

\begin{document}
\maketitle

\begin{abstract}
We show that with a realistic treatment of spectator momentum distributions the RHIC detector trigger sensitivity is high even when RHIC is run below injection energies. In particular, a problem region with $\sqrt{s_{NN}}$=10 to 60 or 80 GeV assuming a simple Fermi step model is not found when using a more realistic one.
We argue also that production of fast nucleons (with momenta $>  p_F$ in the nucleus rest frame) provides complementary information about the collision process. 
\end{abstract}

\section{Introduction}

	In the coming year the RHIC experiments will take data at very low beam energies ($\sqrt{s}\sim 5-50$GeV/u) in order to explore the QCD phase transition critical point. At these energies spectator nucleons from beam breakup are emitted at large angles and must be considered when calculating the trigger efficiency.
	
	Whereas existing heavy ion event generators produce spectators with either no momentum or according to a simple Fermi step distribution, existing data favor a more complicated distribution. We show that for this calculation the above approximations lead to significant errors.

	In an earlier note the PHENIX Zero Degree Calorimeter (ZDC) efficiency at low $\sqrt{s}$ was calculated using Thomas- Fermi(TF) and Feshbach-Huang(FH) (see \cite{Feshbach}) distributions. Feschbach-Huang fits available
data(E814\cite{E814} and also Bevalac) better than Thomas-Fermi. It yields a gaussian momentum distribution with the same r.m.s. value as the Fermi one. 
Of the 2 distributions F-H gave a better low energy efficiency for the ZDC.
We also calculated the acceptance for evaporation neutrons\cite{Evaporation} and found that they dominate the ZDC trigger efficiency.
	
	In this note we discuss the efficiency of the PHENIX Beam-Beam Counters (BBC) (which cover $\eta $=3.1-$>$3.9) due to spectator protons.	
We discuss in some detail the physics of the spectator proton distributions. We also show that a hard component in
the spectrum, usually neglected and due to short range correlations in cold nuclear matter, alters significantly the BBC efficiency.

	This component is not ruled out by E814, for example, because their acceptance is limited to $p_T<$250 MeV. On the other hand it is confirmed by a number of experiments at JLAB, BNL, etc. A popular review can be found in CERN Cour.49N1:22-24,2009. For the experiments see refs.\cite{Tang:2002ww}-\cite{Egiyan:2005hs}.\\
	It also impacts significantly the low beam energy BBC acceptance as can be seen in the following video:
	http://www.phenix.bnl.gov/phenix/WWW/publish/swhite/all3.avi
	
	In the following we first calculate the detection efficiency of the BBC for single spectator protons. We then use NA49 \cite{NA49} measured proton multiplicities to calculate the trigger efficiency of the BBC. \\
\begin{figure}
\centering
\includegraphics[width=0.99\linewidth]{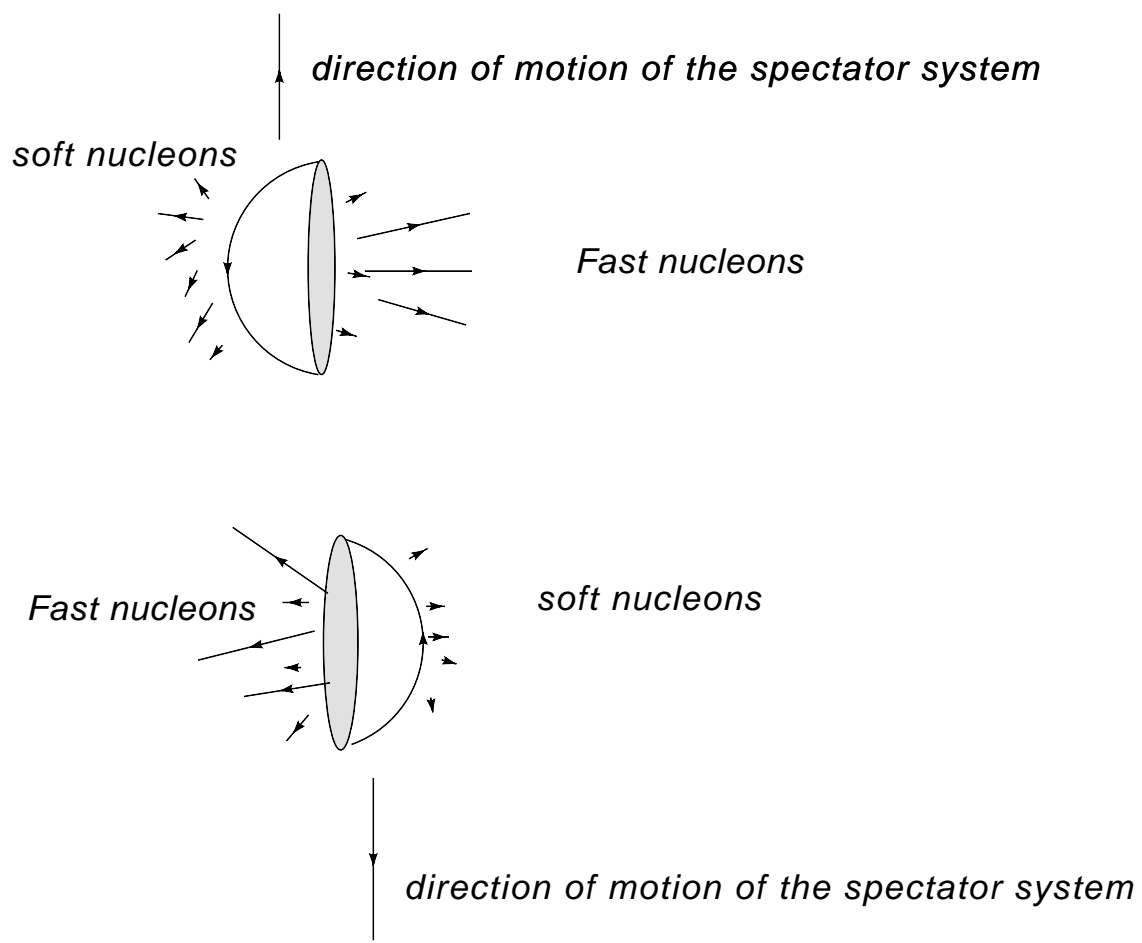}
\caption{Spectator emission.}
\label{fig:Layout}
\end{figure}

\subsection{The Physics of Fragmentation Distributions}

	It's been known for a long time that the scattering of high energy photons, leptons,  hadrons, and nuclei off stationary  nuclei results in significant production of nucleons with momenta $p\ge p_F$ ( $p_F$ is the Fermi momentum) in a wide range of angles including backward angles where fragmentation of wounded  nucleons does not contribute. It was suggested that  the main mechanism of nucleon production is the destruction of short-range correlations in  nuclei
\cite{Frankfurt:1981mk}. This model allows one to describe a number of universal features of these processes including absolute rates. It also led to the prediction  \cite{Frankfurt:1981mk} of 
correlations in semiexclusive reactions like A(e,e'pN) and A(p,2pn) and scaling of the ratios of the (e,e') cross sections at $x> 1.4$.
These predictions were recently confirmed by experiments at the BNL AGS, and Jlab. For a review see \cite{Frankfurt:2008zv}.

In the process of nucleus- nucleus interactions, fragmentation of the nucleus at a given impact parameter, b, should depend only weakly  on the incident energy (the main effect is a change of $\sigma_{inel}(NN)$). However one would expect different dynamics for small and large b.

{\it Small impact parameters}

	For small b most of the nucleons are knocked out and as a result there is a crescent 
shaped volume - practically a two dimensional surface. As a result the average distance between surviving nucleons will be comparable to the internucleon distances and hence the spectator system  would primarily decay into a system of nucleons with momenta similar to those in the initial wave function - that is, the spectator mechanism will be effective.
Hence in this limit we can describe the spectrum approximately as in the spectator mechanism (for simplicity we will give non-relativistic expressions for the cross section). In this case the distribution of produced nucleons over momenta
(in the nucleus rest frame) 
 is
\begin{equation}
\frac{d N}{d^3p} =  (1+p_3/m_N)\, n(p).
\end{equation}
where the $3d$ direction is along the projectile direction. $ (1+p_3/m_N)$  is the flux factor. 

	So there are fewer nucleons emitted backward. $n(p)$ is the momentum distribution which for momenta $p\ge p_0=300 $ MeV/c is dominated by two-nucleon short-range correlations(SRC). The probability of these SRC  is $\sim 25\%$ for heavy nuclei ($\int d^3p \, n(p) \, \theta (p-p_0)$), and 
$n(p) \propto \exp (- \lambda p)$, $\lambda \sim 5.6 $GeV$^{-1}$ , for $p\ge p_0$. One could also use a more realistic wavefunction than discussed there.

	Since the momentum tail in this case is much larger and drops more slowly with $p$, the rate of large angle emission is much stronger in this case than in the Fermi step model.

{\it Medium  impact parameters}

If $b\sim R_A$ the spectator system is roughly a hemisphere.
In this case emission of nucleons due to  the destruction of SRC near the surface of the hemisphere $\sim \pi R_A^2$  leads to emission into the open volume which is given by Eq.(1).  A rough estimate of the number of nucleons near the surface facing the collision volume, given the mean density of nucleons in nuclear media of  $\rho_0\sim 0.17 nucleon/fm^3$, is 
$\pi R_A^2\rho_0 \sim 25 - 30 $ nucleons.  This is  about 1/4 of the spectator nucleons.
The nucleons emitted into the volume of the hemisphere are interacting with the nuclear media and are mostly absorbed  leading to heating of the volume of the nucleus. 

	There is an  additional excitation of the spectator system due to  the creation of holes near the surface  due to the removal of low momentum nucleons.  Since the overall excitation energy is $\sim 40 MeV \times A^{2/3}$ this energy is divided  between  $A/2$ nucleons corresponding to 10 - 15 MeV per nucleon. Hence the spectrum of produced 
nucleons should be quite  soft -  and a significant part of the fragmentation should be into nuclear fragments.  Accordingly, in this case, the production of nucleons  should contain two components - a faster one
predominantly
emitted in the direction of the open hemisphere and a slower mostly isotropic one.
Other mechanisms of fast proton production involve secondary hadron interactions with the nucleus  and elastic/ diffractive  NN near the periphery of one of the nuclei. These contributions do not change qualitatively our conclusions as 
they also lead to a significant  fraction of nucleons being produced with transverse momenta $> p_F$.

\subsection{The momentum distributions}

	We use normalized distributions so the integral over momentum space is always one.

	The Thomas-Fermi distribution is isotropic and uniform in momentum up to  $p<p_F$ for which we use 270 MeV- roughly what is found for heavy nuclei. The mean $p^2$ is =3$p_F^2/5.$

\begin{equation}
\frac{dN^{TF}}{dp_T}=\frac{3p_T \sqrt{p_F^2-p_T^2}}{p_F^3}\cdot \theta(p_F-p_T)
\end{equation}
Feschbach and Huang and also Goldhaber\cite{Goldhaber} derived the observed Gaussian spectra (at the Bevalac) of light fragments. The calculation
is based on clustering in the nucleus. Their distribution yields the same mean - ie $<p^2>$=3$p_F^2$/5  but their distribution
in $p_T$ peaks lower and extends to higher momenta. Although the calculation accounts well for the measured spectra it does less well when compared
with today's picture of the wave function of light 
and heavy
nuclei.

\begin{figure}
\centering
\includegraphics[width=0.99\linewidth]{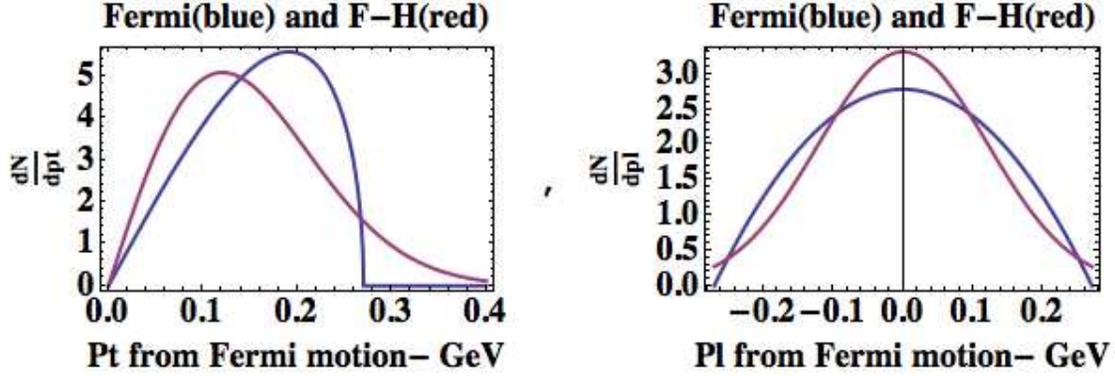}
\caption{The Feshbach-Huang distribution extends to slightly larger $p_L$ and $p_T$ than the Fermi distribution.}
\label{fig:Layout1}
\end{figure}

We now consider the distribution discussed in the two nucleon short-range correlation model by  Frankfurt and Strikman \cite{Frankfurt:1981mk}. This has a dominant component which is Thomas - Fermi like and a
harder component beyond the Fermi momentum. The latter is normalized by the observed 25 $\%$ short range correlation probability (SRC). For our purposes
it is not that significant that the wavefunction for $p<p_F$ is uniform rather than the gaussian of F-H. The critical aspect is the hard component.\\

A simplified version of their wave function which is sufficient for the purposes of our analysis is given by 
\begin{equation} 
\psi ^2 (p) = a\cdot \theta(p_F - p) + b\cdot \exp (-\lambda p)\cdot \theta (p - p_F)
\end{equation}
	
 It enters in Eq.(1)  and leads to the spectator distribution
in the nucleus rest frame which is not isotropic.\\

\begin{figure}
\centering
\includegraphics[width=0.99\linewidth]{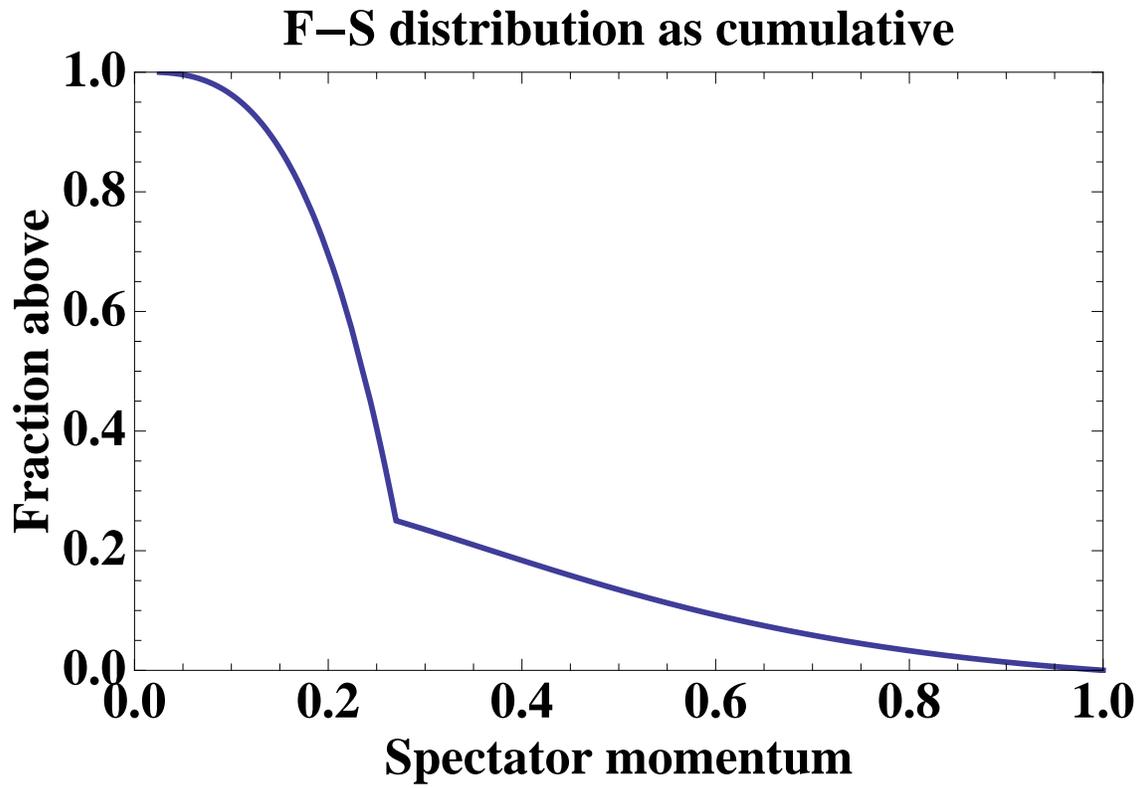}
\caption{The Frankfurt -Strikman distribution as a cumulative.}
\label{fig:Layout2}
\end{figure}

The effect of Fermi smearing on longitudinal momentum is small. Let's evaluate it using a Thomas Fermi distribution. We find an r.m.s. of 120 MeV/c.

Similarly if we define the light cone fraction, $\alpha $, given below, which is boost invariant, we find that $<\alpha >$=1 and $\delta \alpha \simeq
\pm $ .13. Since y$\cong $-ln((E-$p_L$)/mp), the rapidity spread relative to that of the nucleus is ln($\alpha $) and so the spread in rapidity 
due to the longitudinal motion is $\pm$ 0.12 and we will ignore it.
\begin{equation}
\alpha (p_L)=\frac{\sqrt{mp^2+p_L^2}-p_L}{mp}
\end{equation}

	The angular coverage by the PHENIX BBC is effectively 2.3 to 5.2$^o$ or  3.1$<$eta$<$3.9.  The Cerenkov response
is fully efficient above$\sim $1.5 GeV proton momentum. 
	The BBC acceptance is determined by the integration limits on the $p_T$ distribution over this angular range. Naively we wouldn't expect much when $P_{beam}\geq$7 GeV/u since then the BBC coverage falls outside the Fermi surface.

	The ZDC aperture is roughly 2.8 mrad relative to the forward direction.

\begin{figure}
\centering
\includegraphics[width=0.99\linewidth]{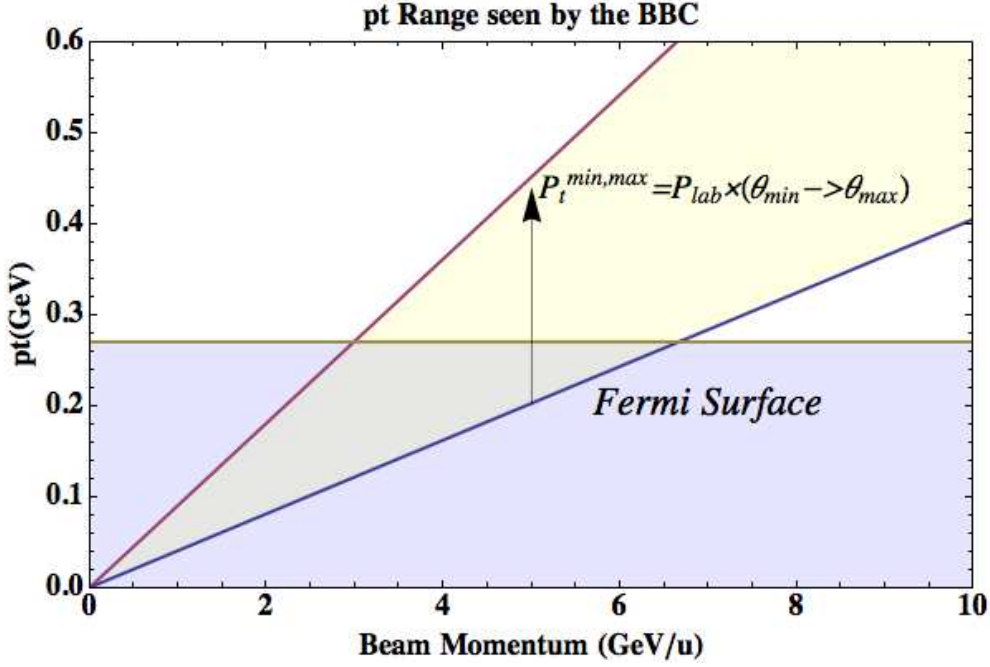}
\caption{ BBC acceptance relative to the Fermi distribution.}
\label{fig:Layout3}
\end{figure}

\begin{equation}
Acceptance(P_{beam})= \int_{P_{beam}\cdot \theta_{min}}^{P_{beam} \cdot\theta_{max}} \frac{dN}{d p_T} d p_T
\end{equation}
\begin{figure}
\centering
\includegraphics[width=0.99\linewidth]{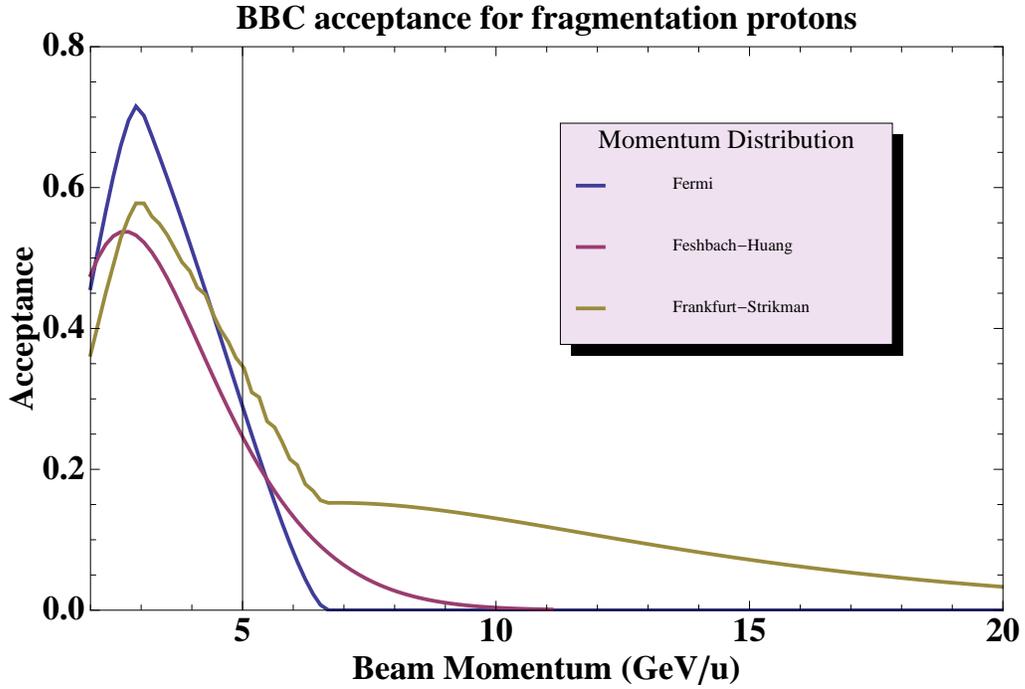}
\caption{Calculated acceptance for one proton in the 3 distributions.}
\label{fig:Layout4}
\end{figure}

We now combine the NA49 measured spectator proton multiplicity with the above 
acceptances to calculate the BBC trigger efficiency due to protons from fragmentation. 

	The NA49 results are presented vs. impact parameter, b, calculated from VENUS and listed here with b(fermi) in the 1st and N(proton) in the 5th column.\\
\begin{figure}
\centering
\includegraphics[width=0.99\linewidth]{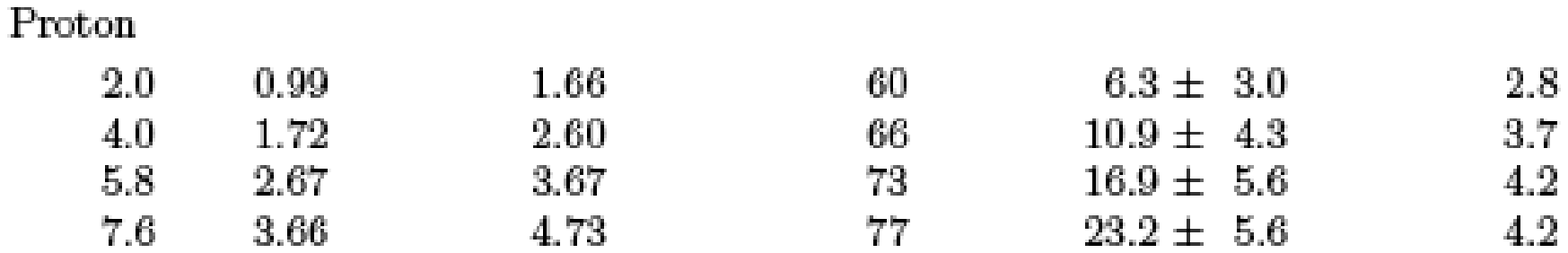}
\caption{NA49's measured spectator proton multiplicity(5) vs calculated b (1).}
\label{fig:Layout5}
\end{figure}

	Below we illustrate the significance of the hard component in BBC trigger efficiency at low energy. The efficiency for a given multiplicity,N,
and acceptance,A, is given by $\varepsilon $=1-(1-A)\(^N\) . For example, with the Fermi distribution and at $p_{beam}$=6 GeV,  A is $\sim $.1 so $\varepsilon
$=0.5 for b=2 and .9 for b=8. Of course $\varepsilon $ is computed for 1 arm.\\

\begin{figure}
\centering
\includegraphics[width=0.99\linewidth]{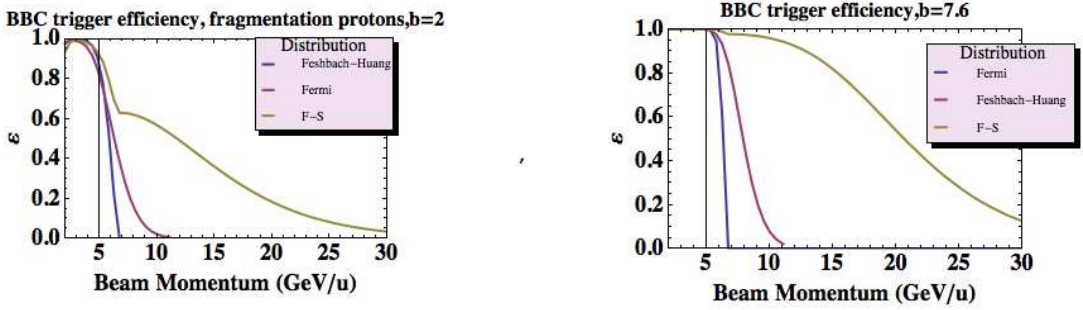}
\caption{BBC trigger efficiency calculated from the NA49 multiplicities.}
\label{fig:Layout6}
\end{figure}

	There is another mechanism for producing spectators at large angles which is usually neglected. Elastic scattering and quasi-elastic(diffractive)
scattering of nucleons will also yield spectators beyond the Fermi surface. We estimate that this is comparable in magnitude to the hard component
from the wave function discussed above. 
In the case of $pA$ collisions a significant contribution to the fast spectator yield comes from the reinteractions of the few GeV hadrons produced in $pA$ scattering. This mechanism is probably less efficient in the case of AA nuclear collisions since production of  hadrons in the nuclear fragmentation region  in AA interactions is likely to be suppressed as compared to the case of pA interactions.

\subsection{Conclusion}

	In conclusion we plot the overall min bias trigger efficiency combining both the ZDC and BBC 2 arm coincidence acceptances. For the ZDC we include the acceptance for evaporation neutrons also. It's interesting that the ZDC and BBC very nearly complement one another and, when combined, give full
coverage over the beam energy range from 2 to 100 GeV. Study of the correlation between the direction of the reaction plane and the direction of the proton emission will provide a novel  probe of the heavy ion collision dynamics  and will also allow one to study it as a function of the incident energy. In particular if the dynamics were to be energy independent the $\eta$ distribution of the spectators would shift with a change of energy simply by $\frac{1}{2} \ln (s_1/s_2)$.

\begin{figure}
\centering
\includegraphics[width=0.99\linewidth]{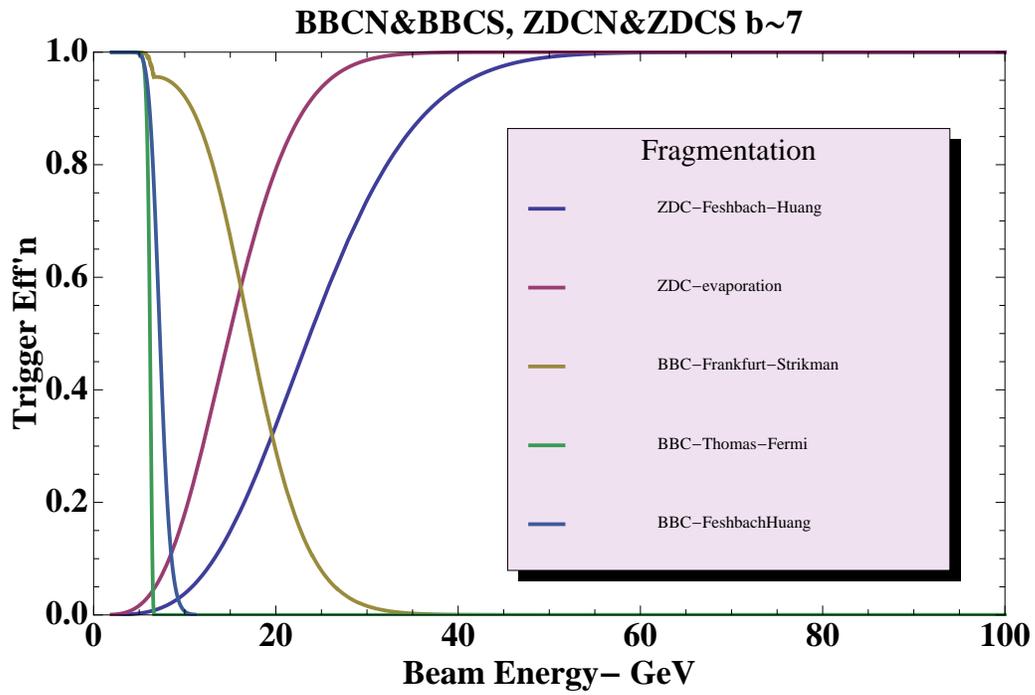}
\caption{Overall trigger Efficiency due to spectator nucleons when requiring either a North arm and South arm coincidence of the BBC or the ZDC.}
\label{fig:Layout7}
\end{figure}

\end{document}